\documentclass[aps,showpacs,twocolumn,prl,superscriptaddress]{revtex4}

\usepackage{amsmath}
\usepackage{latexsym}

\newcommand{\Msun}{\>M_{\odot}}
\newcommand{\beq}{\begin{equation}}
\newcommand{\eeq}{\end{equation}}
\newcommand{\bea}{\begin{eqnarray}}
\newcommand{\eea}{\end{eqnarray}}
\newcommand{\bwt}{\begin{widetext}}
\newcommand{\ewt}{\end{widetext}}
\def\leq{\raise 0.4ex\hbox{$<$}\kern -0.8em\lower 0.62ex\hbox{$-$}}
\def\geq{\raise 0.4ex\hbox{$>$}\kern -0.7em\lower 0.62ex\hbox{$-$}}
\def\lsim{\raise 0.4ex\hbox{$<$}\kern -0.8em\lower 0.62ex\hbox{$\sim$}}
\def\gsim{\raise 0.4ex\hbox{$>$}\kern -0.7em\lower 0.62ex\hbox{$\sim$}}
\def\pm{\,\raise 0.4ex\hbox{$+$}\kern -0.8em\lower 0.62ex\hbox{$-$}\,}

\begin{document}

\title{Constraining the braneworld with gravitational wave observations}
\author{Sean T. McWilliams \\
{\small \emph{Gravitational Astrophysics Laboratory, NASA Goddard Space Flight Center, 8800 Greenbelt Rd., Greenbelt, MD 20771, USA}}}

\date{December 23, 2009}


\begin{abstract}

Some braneworld models may have observable consequences that, if detected, would validate a requisite 
element of string theory. In the infinite Randall-Sundrum model (RS2), the AdS radius of 
curvature, $\ell$, of the extra dimension supports a single bound state of the massless 
graviton on the brane, thereby reproducing Newtonian gravity in the weak-field limit. However, 
using the AdS/CFT correspondence, it has been suggested that one possible consequence of RS2 is 
an enormous increase in Hawking radiation emitted by black holes. We utilize this 
possibility to derive two novel methods for constraining $\ell$ via gravitational wave measurements. 
We show that the EMRI event rate detected by LISA can constrain $\ell$ at the $\sim 1\,\mu$m level 
for optimal cases, while the observation of a single galactic black hole binary with LISA results in 
an optimal constraint of $\ell\,\leq\, 5\mu$m.

\end{abstract}


\maketitle

\paragraph{Introduction.}
String theory requires the existence of extra dimensions.  One variation, the infinite Randall-Sundrum model (hereafter RS2), avoids the necessity of compactifying all but the three observed spatial dimensions, 
by including a bound state of the massless graviton on the brane \cite{RS2} resulting from the curvature,
rather than the size, of the extra dimension.  Since the standard model fields are constrained to the brane by construction, RS2 satisfies all current observational constraints.  However, by
employing the AdS/CFT correspondence \cite{Malda}, it has been predicted that RS2 would result in a substantial enhancement in the amount of Hawking radiation from black holes \cite{Emparan02, Tanaka} due to 
an enormous increase in the number of accessible CFT modes.  A counter-example was derived in \cite{Fitz06} in the form of a static black hole solution, but that solution is unstable.  In the absence of a dynamically stable
solution, there remains the possibility that the RS2 model results in a tremendous increase in the amount
of Hawking radiation emitted by macroscopic black holes.
Several authors have therefore employed the results of \cite{Emparan02, Tanaka}, particularly the modified mass-loss rate due to Hawking radiation, to place constraints on the asymptotic AdS radius of curvature, $\ell$.

RS2 predicts a modification of the Newtonian potential at second post-Newtonian order (2PN),
\beq
V=-\frac{GMm}{r}\left[1+\frac{2}{3}\left(\frac{\ell}{GM}\right)^2\left(\frac{GM}{r}\right)^2+\ldots\right]\,,
\label{eq:vnewt}
\eeq
where, throughout this work, we will use $m$ for the stellar black hole mass, and $M$ for the mass of the companion,
be it a white dwarf, neutron star, or black hole.
Table-top measurements of gravity on the ${\mathcal O}(10\,\mu$m) scale have placed strict upper limits on $\ell$ (44 $\mu$m in \cite{Kapner07}, 
15 $\mu$m in \cite{Geraci08}).  Constraints on $\ell$ have also been derived from astronomical
observations \cite{Emparan03, Psaltis07, Johan09a, Johan09b, Psaltis09} and from the potential observation of evaporating primordial
black holes (see \cite{Kavic} and references therein).
The impact of large extra dimensions on gravitational waves has been studied to an extent, though primarily in a cosmological context (see \cite{Seahra},
 as well as \cite{Maartens} and references therein).
However, while in braneworld models with two branes, solutions such as large black strings are known to be dynamically stable, no such solution has been 
found in the effectively-single brane RS2 model.  This makes
it difficult to directly calculate the effect of $\ell$ on the gravitational radiation emitted on the brane in RS2 self-consistently.  

One proposal to constrain RS2 with gravitational waves involved the observation of very high frequency gravitational
waves from sub-lunar mass black holes \cite{Lunar}.  That work did not take into account any enhancement in Hawking radiation, 
and therefore only investigated the measurability of $\ell$ through its modification of the Newtonian potential
at 2PN order assuming a constant mass.  We also note that
$M\rightarrow M(t)$ can be reinterpreted as $G\rightarrow G(t)$, as $G$ and $M$ appear together everywhere
in the Hamiltonian with the
exception of the rest mass of the black hole, so that methods to constrain $\dot{G}$ with gravitational waves 
\cite{Yunes} may also yield useful constraints
on $\ell$ and vice versa.

In this Letter, we find that gravitational wave measurements 
involving stellar black holes ($2\Msun\leq m\,\lsim\, 20\Msun$) with the proposed Laser Interferometer Space Antenna
(LISA) can place the strongest limits on $\ell$ 
of any proposed mechanism not involving black holes less massive than the Tolman-Oppenheimer-Volkoff limit.
We derive two constraints, one from the event rate of stellar black holes inspiraling gravitationally
into supermassive black holes ($M\simeq 10^6\Msun$), and the second from the observation of individual galactic binaries
containing a stellar mass black hole (GBHBs hereafter).
The first type of system, referred to as an extreme mass ratio inspiral (EMRI),
is expected to be a primary source of detectable signals for LISA, with thousands of events during the mission lifetime \cite{Gair}.  However,
since enhanced Hawking radiation would cause stellar black holes to evaporate, it would significantly impact the number of detectable EMRIs, so we may
take advantage of the expectation of a diminished event rate to constrain $\ell$.
The second type of system, GBHBs, will be far less numerous than galactic white dwarf-white dwarf binaries, but population synthesis simulations 
have predicted the detection of
$\sim 10$ GBHB systems over the LISA lifetime \cite{Nelemans}.  None of these systems in the referenced simulation 
were black hole-black hole binaries, and such systems
are poorly constrained observationally.  However, other simulations (e.~g.~\cite{Voss}) predict them to be vastly more
common than other varieties of GBHB.  Should they be detected by LISA, double black hole binaries 
would yield a comparable constraint to what we find for GBHBs.  Hawking radiation, as we will show, has the effect of driving a GBHB to outspiral,
thereby preventing it from evolving into the LISA band.  Therefore, by observing a GBHB in the LISA band, we can derive a constraint on $\ell$.
We find that, among constraints deriving from stellar mass black holes, both methods
presented in this Letter result in limits that are comparable to the most stringent in the literature.  Furthermore,
while the constraint derived from EMRI rates depends on a number of assumptions and has significant uncertainty, the constraint derived from
observing GBHBs has no such dependence, and should provide a robust limit for the AdS curvature
in RS2.

\paragraph{Constraining $\ell$ with EMRI event rates.}
For both of the scenarios we present for constraining the RS2 model, we compare the observational expectations
derived by assuming standard (3+1) general relativity with the modified expectations resulting from enhanced Hawking
radiation in the RS2 model.  
We first focus on EMRI systems, consisting of a stellar mass black hole merging with a supermassive
black hole.  The principle formation mechanism for EMRIs is the direct capture of a stellar black hole by the supermassive black hole
at a galactic center, after the stellar black hole has undergone large angle scattering \cite{Sigurdsson}.
If a stellar black hole passes within a critical distance of a supermassive black hole, it will emit enough energy in gravitational waves as it 
passes to become bound.  At that point, it will inspiral deterministically
due to the loss of angular momentum from gravitational radiation.  
Because enhanced Hawking radiation will shorten the lifetime of black holes that could otherwise potentially be captured by
a supermassive black hole, we can derive a constraint on $\ell$ from the event rate of EMRIs
formed by direct capture.
From \cite{Gair}, the average event rate for EMRIs is
\beq
\langle{\mathcal R}\rangle_{\rm EMRI} \approx \left(\frac{M}{10^6\Msun}\right)^{\frac{3}{8}}\sqrt{\frac{5\Msun}{m}}\, {\rm Gpc^{-3}\,yr^{-1}}\,,
\label{eq:emrate}
\eeq
where events were binned by dividing $M$ into fixed logarithmic intervals of $0.5$.
This estimate assumes that the average timescale for a compact object to undergo gravitational capture
by the central black hole is approximately half the dynamical friction timescale.
By calculating the number of compact objects that will be captured in less than the galactic age, which they take to be
$10^{10}$ yr, they predict an event rate for EMRIs.
The direct capture timescale used in \cite{Gair} is given by
\beq
t_{\rm dc}\approx \frac{t_{\rm df}}{2} \approx 0.15\frac{\sigma^3}{G^2m\rho_*\ln \Lambda}\,{\rm yr} 
=0.15\frac{2\pi r^2}{Gm\ln \Lambda}\,{\rm yr}\,,
\label{eq:tdf}
\eeq
where $t_{\rm df}$ is the dynamical friction timescale, $\rho_*$ is the local stellar density, 
given by $\sigma^2/(2\pi G r^2)$, $\sigma$ is the spheroid velocity dispersion, and $\ln \Lambda = 5$
is a parameter incorporating the range of potential impact parameters.  In \cite{Gair}, the authors calculate an effective
distance from the center, $r_{\rm dc}$, inside which compact objects will be captured, by requiring that $t_{\rm dc}$ not exceed $10^{10}$ yrs.
However, if enhanced Hawking mass loss were to occur, at least some systems that would otherwise merge with the central supermassive black hole
within $10^{10}$ yrs will instead evaporate due to Hawking radiation.
The evaporation timescale for a stellar black hole in RS2, assuming the AdS/CFT correspondence, is given by \cite{Emparan03}
\beq
t_{\rm H}=7.7\times 10^6 \left(\frac{m}{5\Msun}\right)^3\left(\frac{44\,\mu{\rm m}}{\ell}\right)^2\,{\rm yr}\,.
\label{eq:life}
\eeq
It is straightforward to compare Eq.~\eqref{eq:tdf} evaluated for $t_{\rm dc}\,\leq\,10^{10}$ yrs and evaluated for $t_{\rm dc}\,\leq\,t_{\rm H}$,
in order to find the relationship between $r_{\rm dc}$ and the modified effective distance under Hawking radiation, which we call $r_{\rm H}$: 
\beq
\frac{t_{\rm H}}{10^{10}\,{\rm yrs}} =\left(\frac{r_{\rm H}}{r_{\rm dc}}\right)^2 = 7.7\times 10^{-4}\left(\frac{m}{5\Msun}\right)^3\left(\frac{44\,\mu{\rm m}}{\ell}\right)^2\,.
\label{eq:rh}
\eeq 
Using Eq.~\eqref{eq:rh}, we can then express the modified average event rate, $\langle{\mathcal R}\rangle_{\rm H}$, as
\bea
\langle{\mathcal R}\rangle_{\rm H} &=& \langle{\mathcal R}\rangle_{\rm EMRI} \times \frac{\int\limits_0^{r_{\rm H}} 4\pi\rho_*r^3/3\,dr}{\int\limits_0^{r_{\rm df}} 4\pi\rho_*r^3/3\,dr} = \left(\frac{r_{\rm H}}{r_{\rm df}}\right)^2\langle{\mathcal R}\rangle_{\rm EMRI} \nonumber \\
&=& 7.7\times 10^{-4}\left(\frac{44\,\mu{\rm m}}{\ell}\right)^2\left(\frac{M}{10^6\Msun}\right)^{\frac{3}{8}} \nonumber \\
&\times& \left(\frac{m}{5\Msun}\right)^{\frac{5}{2}}\,{\rm Gpc^{-3}\,yr^{-1}}\,. 
\label{eq:rateh}
\eea
The distribution of event rates in a given volume will follow a Poisson distribution, so that the variance
will equal the mean.
We therefore find a standard deviation of 
\bea
\sigma_{\mathcal R} &=& 0.028 \left(\frac{44\,\mu{\rm m}}{\ell}\right)\left(\frac{M}{10^6\Msun}\right)^{\frac{3}{16}} \nonumber \\
&\times& \left(\frac{m}{5\Msun}\right)^{\frac{5}{4}} \,{\rm Gpc^{-3}\,yr^{-1}}\,.
\label{eq:sdntot}
\eea
Using Eq.~\eqref{eq:sdntot}, we can therefore derive the $5$-$\sigma$ limit
on $\ell$ in the scenario that the observed event rate is given by Eq.~\eqref{eq:emrate}. 
We find a constraint
given by
\beq
\ell_{5\sigma} \,\lsim \, 6.1 \left(\frac{m}{5\Msun}\right)^{\frac{7}{4}}\left(\frac{10^6\Msun}{M}\right)^{\frac{3}{16}}\,\mu{\rm m}\,.
\label{eq:lemri}
\eeq
Assuming a minimum black hole mass of $2\Msun$, Eq.~\eqref{eq:lemri} predicts an optimal constraint of $\sim 1\,\mu$m.
This constraint is therefore comparable to
the most stringent constraint from stellar black holes in the literature ($3\,\mu$m in\cite{Psaltis09}), which
was derived by combining the measurement of a black hole's mass with an estimation of the age of its cluster.
However, like that limit, the precise constraint derived from the EMRI event rate depends on auxiliary assumptions.
Specifically, our constraint is limited by the astrophysical uncertainties in the nominal EMRI event rate, such as the density
profile of stars in the center of the host galaxy, the age of the host galaxy, and the timescale for direct capture.
We do note, however, that the ratio 
$\langle{\mathcal R}\rangle_{\rm H}/\langle{\mathcal R}\rangle_{\rm EMRI}$ should be relatively robust,
so improved knowledge of the astrophysics that determines the nominal EMRI event rate will likewise increase our confidence in the
constraint.  In the next section, we derive a constraint from GBHBs that, while somewhat less stringent, should be quite
robust, as it relies on the observation of a single event, and requires no secondary assumptions. 

\paragraph{Constraining $\ell$ through the observation of galactic binaries.}
GBHBs are relatively weak gravitational wave emitters, with separations of ${\mathcal O}(10^{-3}$ AU), or ${\mathcal O}(10^{4}\, M)$
for the masses of interest.  Their signals are therefore nearly monochromatic
during the LISA mission lifetime.  GBHBs accumulate enough SNR to be detectable both above detector noise,
as well as above the confusion noise resulting from the density in frequency space of galactic white dwarf-white dwarf binaries, due
in part to the larger mass of GBHBs \cite{Nelemans}.  In fact, since binaries spend much more time at low frequencies than high,
the simulations in \cite{Nelemans} indicate that nearly all detectable GBHBs occur at confusion-limited frequencies of $\sim 0.1$ mHz, for
black hole masses of $\sim 5\Msun$, so we will take these as fiducial values in the analysis that follows.  At such low frequencies,
GBHBs will not have measurable chirps \cite{Seto}, but in the cases of black hole-black hole, black hole-neutron star, and 
detached black hole-white dwarf binaries, the systems will have some measurable eccentricity due to the supernova kick of one or both
components when they enter the LISA band \cite{Voss} (semi-detached black hole-white dwarf systems will circularize due to mass transfer \cite{vanHaam}).
This eccentricity will
make it possible to measure the total mass of the system with a fractional error of a few percent \cite{Seto2}, 
which can be used to infer that a system is 
sufficiently massive to necessitate it being a GBHB.  

We assume a neutron star mass of $2\Msun$, and therefore assume that a total mass
of $7\Msun$ has been measured.  Clearly, the measurement of such a large mass indicates that one member is a black hole.  In fact,
both members could be black holes, but assuming only one would be the conservative approach, since two black holes of smaller mass
would be more affected by Hawking radiation at the same $\ell$, and would therefore imply a tighter constraint on $\ell$.  Assuming
the GBHB consists of a black hole and a white dwarf would be slightly more conservative, so we wish to include the appropriate
scaling relationships in our constraint, so that we can apply it for different component mass combinations for a fixed total mass..

One might object that, without observing an inspiral chirp, we cannot be certain that
the system is not outspiralling very slowly under the influence of Hawking radiation, so we will address this possibility
before proceeding.  GBHBs for the masses of
interest will have orbital periods of ${\mathcal O}($days) immediately after the GBHB is formed from the stellar binary progenitor \cite{Kalogera}, 
and will therefore be slightly below the LISA sensitivity band at formation.
GBHBs must then inspiral gravitationally in order to be detectable by LISA.  Since any system that is initially Hawking or
gravitational radiation-dominated will remain so, the observation of a GBHB in the LISA band requires that the GBHB has inspiralled
to that frequency gravitationally, and has therefore always been gravitational radiation-dominated.  Therefore, the assumption
that any detected GBHB is gravitational radiation-dominated is justified.

As we have mentioned, gravitational radiation will drive GBHBs to slowly inspiral,
while Hawking radiation will drive the system to slowly outspiral.  To see this, we note that since Hawking radiation is isotropic, 
the specific angular momentum $j$ will be conserved, since $\dot{j}=\dot{r}v_{\phi}+r\dot{v}_{\phi}=0$.
We can therefore use conservation of $j$ to calculate the derivative of the semi-major axis, $a$, under the influence of Hawking radiation.
Differentiating $j=\sqrt{Ga(M+m)}$ yields
\bea
\dot{a}_{\rm H}&=&-\frac{a\frac{d}{dt}(M+m)}{M+m}=-\frac{a\dot{m}_{\rm H}}{M+m} = 3.2\times 10^{-8}\left(\frac{a}{1\,{\rm AU}}\right) \nonumber \\
&\times& \left(\frac{7\Msun}{M+m}\right)\left(\frac{5\Msun}{m}\right)^2\left(\frac{\ell}{44\,\mu{\rm m}}\right)^2\,{\rm AU\,yr^{-1}}\,,
\label{eq:adothawk}
\eea
where $\dot{m}_{\rm H}$ is the mass loss due to Hawking radiation of massive gravitons into the bulk
\citep{Tanaka, Emparan02},
\beq
\dot{m}_{\rm H}=-2.2\times 10^{-7}\left(\frac{5\Msun}{m}\right)^2\left(\frac{\ell}{44\,\mu{\rm m}}\right)^2\,\Msun\,{\rm yr^{-1}}\,.
\label{eq:mdoth}
\eeq
We note that, to be consistent with the preceding section, $m$ represents the mass of the stellar black hole, although it is now more
massive than its compact companion, for which we use a fiducial $M=2\Msun$ for the mass of a neutron star.
We emphasize that Eq.~\ref{eq:adothawk} is positive, meaning the mass loss from Hawking radiation
drives the GBHB to outspiral.  The rate of inspiral due to gravitational radiation is given, at Newtonian order, by
\bea
\dot{a}_{\rm gw}&=&-\frac{64G^3Mm(M+m)}{5c^5a^3} = -5.2\times 10^{-17}\left(\frac{1\,{\rm AU}}{a}\right)^3 \nonumber \\
&\times& \left(\frac{m}{5\Msun}\right)\left(\frac{M}{2\Msun}\right)\left(\frac{M+m}{7\Msun}\right)\,{\rm AU\,yr^{-1}}\,.
\label{eq:adotgw}
\eea
Equating Eqs.~\ref{eq:adothawk} and \ref{eq:adotgw} and solving for the semi-major axis yields the
critical separation, $a_{\rm crit}$, separating the Hawking- and gravitational radiation-dominated regimes:
\bea
a_{\rm crit}&=&6.3\times 10^{-3}\left(\frac{m}{5\Msun}\right)^{\frac{3}{4}}\left(\frac{M}{2\Msun}\right)^{\frac{1}{4}} \nonumber \\
&\times& \sqrt{\left(\frac{M+m}{7\Msun}\right)\left(\frac{44\,\mu{\rm m}}{\ell}\right)}\,{\rm AU}\,.
\label{eq:acrit}
\eea

In the event that a gravitational signal is detected at a frequency $f_{\rm gw}$, we can use Kepler's law 
(and consider only quadrupolar radiation, so $f_{\rm gw}=2f_{\rm orb}$) to find the corresponding semi-major axis,
\bea
a &=& \left(\frac{G(M+m)}{(\pi f_{\rm GW})^2}\right)^{\frac{1}{3}} \nonumber \\
&=& 8.9\times 10^{-3}\left(\frac{M+m}{7\Msun}\right)^{\frac{1}{3}}\left(\frac{0.1\,{\rm mHz}}{f_{\rm GW}}\right)^{\frac{2}{3}}\,{\rm AU}\,.
\label{eq:agb}
\eea
where we use a fiducial frequency of 0.1 mHz, which 
we again note is informed by the results for GBHBs in \cite{Nelemans}.
By requiring that the semi-major axis from Eq.~\eqref{eq:agb} be less than that from 
Eq.~\eqref{eq:acrit}, we find a constraint on $\ell$ of
\beq
\ell\,\leq\,22\left(\frac{m}{5\Msun}\right)^{\frac{3}{2}}\left(\frac{M+m}{7\Msun}\right)^{\frac{1}{3}}\sqrt{\frac{M}{2\Msun}}\left(\frac{f_{\rm gw}}{0.1\,{\rm mHz}}\right)^{\frac{4}{3}}\,\mu{\rm m}\,.
\label{eq:lgb}
\eeq
As this constraint derives from a direct
measurement of a single event that does not depend on auxiliary assumptions, it should provide a robust
limit on the asymptotic radius of curvature of an infinite extra dimension.  If we again employ a minimum black hole mass
of $2\Msun$, Eq.~\eqref{eq:lgb} implies an optimal constraint of $5\,\mu$m, although it would be difficult to distinguish
such a system from a double neutron star binary, which would not be subject to enhanced Hawking radiation.
Our GBHB constraint is analogous to \cite{Johan09a} in that both derive a
constraint from the orbital dynamics of a stellar black hole in a binary system, and both depend minimally on auxiliary assumptions.  
Our methods are therefore
equivalently robust, with our result being somewhat more stringent than the $35\,\mu$m constraint on $\ell$ found in \cite{Johan09a}.

\paragraph{Conclusions.}
For stellar-mass black holes, we find
a constraint on $\ell$ using two separate methods, the first involving the observation of EMRIs with LISA, and 
the second involving the observation of GBHBs.  The latter method provides our most
robust constraint of $\ell\,\leq\,22\,\mu$m for typical cases and $\ell\,\leq\,5\,\mu$m for optimal cases,
which is effectively equivalent to the limit set by estimating
the age of a stellar black hole in a globular cluster \cite{Psaltis09} and improves on the limit from x-ray binaries \cite{Johan09a}.  
The optimal constraint of $1\,\mu$m derived from the EMRI event rate exceeds that from GBHBs, 
although it is a less robust limit.

\paragraph{Acknowledgments.}
We wish to thank John Baker, Ira Thorpe, and Tim Johannsen for useful discussions,
and Cole Miller, Frans Pretorius, and Nico Yunes for helpful comments on this manuscript.

\end{document}